\documentclass[twoside,slac_one]{revtex4}
\usepackage{graphicx}
\usepackage{fancyhdr}
\usepackage{amsmath} 
\usepackage{bm}
\usepackage{amsxtra}
\usepackage{amssymb}
\usepackage{amsthm}
\usepackage{latexsym}
\usepackage{lscape}
\usepackage{subfigure}
\usepackage{atlasphysics}
\usepackage{booktabs} 
\usepackage{epstopdf}
\usepackage{hyperref}

\def\lumi{1.02}

\def\lumiRelErr{3.7} 

\def\WZtotXsecExpected{17.2}
\def\WZtotXsecExpectedErrUp{1.2}
\def\WZtotXsecExpectedErrDw{-0.8}

\def\WZtotXsec{21.1}
\def\WZtotXsecStatErrUp{3.1}
\def\WZtotXsecStatErrDw{2.8}
\def\WZtotXsecSysErrUp{1.2}
\def\WZtotXsecSysErrDw{1.2}
\def\WZtotXsecLumiErrUp{0.9}
\def\WZtotXsecLumiErrDw{0.8}
\def\WZfidXsec{118}
\def\WZfidXsecStatErrUp{18}
\def\WZfidXsecStatErrDw{16}
\def\WZfidXsecSysErrUp{6}
\def\WZfidXsecSysErrDw{6}
\def\WZfidXsecLumiErrUp{5}
\def\WZfidXsecLumiErrDw{5}

\def\WZNSigEventsExpected{49.1}
\def\WZNSigEventsExpectedErrStat{0.4}

\def\WZNSigEventsExpectedErrSys{3.02}

\def\WZNSigEventsObserved{71}
\def\WZNSigEventsExpectedEEE{7.55}
\def\WZNSigEventsExpectedErrStatEEE{0.17}

\def\WZNSigEventsObservedEEE{11}
\def\WZNSigEventsExpectedEEM{11.27}
\def\WZNSigEventsExpectedErrStatEEM{0.20}

\def\WZNSigEventsObservedEEM{9}
\def\WZNSigEventsExpectedEMM{12.12}
\def\WZNSigEventsExpectedErrStatEMM{0.22}

\def\WZNSigEventsObservedEMM{22}
\def\WZNSigEventsExpectedMMM{18.16}
\def\WZNSigEventsExpectedErrStatMMM{0.27}

\def\WZNSigEventsObservedMMM{29}
\def\WZSB{4.3}
\def\WZNBkgEventsExpected{10.5}

\def\WZNBkgEventsExpectedErrStat{0.8}

\def\WZNBkgEventsExpectedErrUpSys{2.9}
\def\WZNBkgEventsExpectedErrDwSys{2.1}
\def\WZNZZBkgEventsExpected{3.55}
\def\WZNZZBkgEventsExpectedErrStat{0.24}
\def\WZNZZBkgEventsExpectedErrSys{0.17}
\def\WZNTopBkgEventsExpected{1.58}
\def\WZNTopBkgEventsExpectedErrStat{0.23}
\def\WZNTopBkgEventsExpectedErrSys{0.10}
\def\WZNZgammaBkgEventsExpected{1.05}
\def\WZNZgammaBkgEventsExpectedErrStat{0.48}
\def\WZNZgammaBkgEventsExpectedErrSys{0.08}
\def\WZNZjetsDDBkgEventsExpected{5.14}
\def\WZNZjetsDDBkgEventsExpectedErrStat{0.59}
\def\WZNZjetsDDBkgEventsExpectedErrSysUp{2.97}
\def\WZNZjetsDDBkgEventsExpectedErrSysDown{2.08}

\def\WZtau{1.7}

\def\WZSBEEE{2.5}
\def\WZNBkgEventsExpectedEEE{3.08}
\def\WZNBkgEventsExpectedErrStatEEE{0.49}

\def\WZNTopBkgEventsExpectedEEE{0.26}
\def\WZNTopBkgEventsExpectedErrStatEEE{0.10}

\def\WZNZZBkgEventsExpectedEEE{0.34}
\def\WZNZZBkgEventsExpectedErrStatEEE{0.07}

\def\WZNZgammaBkgEventsExpectedEEE{0.49}
\def\WZNZgammaBkgEventsExpectedErrStatEEE{0.28}

\def\WZNZjetsDDBkgEventsExpectedEEE{2.03}
\def\WZNZjetsDDBkgEventsExpectedErrStatEEE{0.38}

\def\WZSBEEM{5.7}
\def\WZNBkgEventsExpectedEEM{1.98}
\def\WZNBkgEventsExpectedErrStatEEM{0.24}

\def\WZNTopBkgEventsExpectedEEM{0.31}
\def\WZNTopBkgEventsExpectedErrStatEEM{0.09}

\def\WZNZZBkgEventsExpectedEEM{1.03}
\def\WZNZZBkgEventsExpectedErrStatEEM{0.13}

\def\WZNZjetsDDBkgEventsExpectedEEM{0.64}
\def\WZNZjetsDDBkgEventsExpectedErrStatEEM{0.18}

\def\WZSBEMM{3.2}
\def\WZNBkgEventsExpectedEMM{3.82}
\def\WZNBkgEventsExpectedErrStatEMM{0.56}

\def\WZNTopBkgEventsExpectedEMM{0.41}
\def\WZNTopBkgEventsExpectedErrStatEMM{0.12}

\def\WZNZZBkgEventsExpectedEMM{0.82}
\def\WZNZZBkgEventsExpectedErrStatEMM{0.12}

\def\WZNZgammaBkgEventsExpectedEMM{0.56}
\def\WZNZgammaBkgEventsExpectedErrStatEMM{0.39}

\def\WZNZjetsDDBkgEventsExpectedEMM{2.03}
\def\WZNZjetsDDBkgEventsExpectedErrStatEMM{0.38}

\def\WZSBMMM{7.4}
\def\WZNBkgEventsExpectedMMM{2.44}
\def\WZNBkgEventsExpectedErrStatMMM{0.21}

\def\WZNTopBkgEventsExpectedMMM{0.60}
\def\WZNTopBkgEventsExpectedErrStatMMM{0.15}

\def\WZNZZBkgEventsExpectedMMM{1.40}
\def\WZNZZBkgEventsExpectedErrStatMMM{0.15}

\def\WZNZjetsDDBkgEventsExpectedMMM{0.44}
\def\WZNZjetsDDBkgEventsExpectedErrStatMMM{0.15}

\def\expglow{-0.16}
\def\expgup{0.24}
\def\expkappalow{-0.7}
\def\expkappaup{0.9}
\def\explambdalow{-0.14}
\def\explambdaup{0.14}
\def\errlowexpglow{0.05}
\def\errupexpglow{0.05}
\def\errlowexpgup{0.05}
\def\errupexpgup{0.05}
\def\errlowexpkappalow{0.2}
\def\errupexpkappalow{0.2}
\def\errlowexpkappaup{0.2}
\def\errupexpkappaup{0.2}
\def\errlowexplambdalow{0.03}
\def\errupexplambdalow{0.04}
\def\errlowexplambdaup{0.04}
\def\errupexplambdaup{0.03}
\def\obsglow{-0.21}
\def\obsgup{0.30}
\def\obskappalow{-0.9}
\def\obskappaup{1.2}
\def\obslambdalow{-0.18}
\def\obslambdaup{0.18}
\def\obsglowSixtyEightA{-0.17}
\def\obsgupSixtyEightA{-0.05}
\def\obskappalowSixtyEightA{-0.8}
\def\obskappaupSixtyEightA{-0.2}
\def\obslambdalowSixtyEightA{-0.15}
\def\obslambdaupSixtyEightA{-0.06}

\def\obsglowSixtyEightB{0.13}
\def\obsgupSixtyEightB{0.26}
\def\obskappalowSixtyEightB{0.5}
\def\obskappaupSixtyEightB{1.0}
\def\obslambdalowSixtyEightB{0.06}
\def\obslambdaupSixtyEightB{0.15}

\begin{document}

\title{Measurement of the \WZ Production Cross Section in Proton-Proton Collisions at $\mathbf{\sqrt{s} =7}\TeV$ with the ATLAS Detector}

%

\author{Yusheng Wu, on behalf of ATLAS collaboration}
\affiliation{University of Michigan, University of Science and Technology of China}

\begin{abstract}
This document presents a measurement of \WZ production in  \lumi~\ifb\ of $pp$ collision data
at $\sqrt s = 7\TeV~$ collected by the ATLAS experiment at LHC in 2011.
A total of \WZNSigEventsObserved\ candidates with a background expectation of 
\WZNBkgEventsExpected$\pm$\WZNBkgEventsExpectedErrStat(stat)$^{+\WZNBkgEventsExpectedErrUpSys}_{-\WZNBkgEventsExpectedErrDwSys}$(sys) 
events were observed for purely leptonically decaying bosons with electrons, muons and missing transverse energy in the final state.
The total cross section has been determined to be 
$\sigma_{WZ}^{tot}=  \WZtotXsec^{+\WZtotXsecStatErrUp}_{-\WZtotXsecStatErrDw}$(stat) $^{+\WZtotXsecSysErrUp}_{-\WZtotXsecSysErrDw}$(syst) $^{+\WZtotXsecLumiErrUp}_{-\WZtotXsecLumiErrDw}$(lumi)~\pb, 
in agreement with the Standard Model expectation of $\WZtotXsecExpected^{+\WZtotXsecExpectedErrUp}_{\WZtotXsecExpectedErrDw}$~\pb. 
Limits on anomalous triple gauge boson couplings have been derived.
\end{abstract}

\maketitle

\thispagestyle{fancy}



\section{Introduction}\label{sec:Introduction}
WZ Production in the standard model (SM) contains triple gauge coupling (TGC) vertex, which is essential to test the high energy behavior of the SM and is also a probe to new physics.
Furthermore, new particles decaying into \WZ pairs are predicted in supersymmetric models with an extended Higgs sector (charged Higgs) as well as models with extra vector bosons (e.g. $W^{\prime}$)~\cite{DKS:1999}.

At the LHC, the dominant \WZ production mechanism is from quark-antiquark and quark-gluon initial states at leading-order (LO) and at next-to-leading order (NLO), respectively~\cite{PhysRevD.65.094041}. Only the $s$-channel diagram has a triple gauge boson interaction vertex and is hence the only channel to contribute to potential anomalous coupling behaviour of gauge bosons.

This note presents a measurement of the \WZ production cross section with the ATLAS detector in LHC proton-proton collisions at $\sqrt{s}$ = 7 TeV using \lumi~\ifb\ of data collected in 2011. 
The analysis uses four channels with leptonic decays ($\WZ\to\ell\nu\ell\ell$) involving electrons and muons: $eee$, $ee\mu$, $\mu\mu e$ or $\mu\mu\mu$ plus missing transverse energy, $\MET$.
The main sources of background to the leptonic \WZ signal are $ZZ$, $Z\gamma$, $Z$+jets, and top-quark events.  
The signal and background contributions are modelled with Monte Carlo simulation and with data-driven measurements.

Section~\ref{sec:ATLAS} briefly describes the ATLAS detector and the data sample analysed in this paper.  
Section~\ref{sec:MCSamples} discusses the signal and background simulation samples used in this analysis. The definition and reconstruction of physical observable objects such as particles and jets are detailed in Section~\ref{sec:Reconstruction}, followed by event selection in Section~\ref{sec:EventSelection}. 
Section~\ref{sec:SignalAcceptance} presents the signal acceptance and background estimate, and the systematic uncertainties on these estimates. The calculations of the cross section and limits on the anomalous TGCs (aTGCs) are given in Section~\ref{sec:Results}.

\section{The ATLAS Detector and Data Sample}\label{sec:ATLAS}
The ATLAS detector is a multipurpose particle physics apparatus including a precision tracking system as the innermost part of the detector, highly segmented electromagnetic and hadronic calorimeters, and a large muon spectrometer~\cite{atlas}.

This study uses data collected between March and June 2011.
Data periods flagged with data quality problems that affect the reconstructed objects used in this analysis are removed.
After data quality cuts, the total integrated luminosity used in this analysis is \lumi~\ifb.
The preliminary luminosity uncertainty for 2011 data is \lumiRelErr\%.


$\WZ$ candidate events with multi-lepton final states are selected online 
with single-muon or single-electron 
triggers requiring \pT\ of at least 18 (20) GeV for muons (electrons). 
The trigger efficiency for \WZ events,
which pass all selection criteria, is in the range of 96--99\%
depending on the final state considered.

\section{Simulated Data Samples}\label{sec:MCSamples}
The \WZ production processes and subsequent pure leptonic decays
are modelled by the \mcatnlo~\cite{Frixione:2002ik} generator, which incorporates the NLO QCD matrix elements into the parton 
shower by interfacing to the~\herwig~\cite{Herwig} program. The parton density function (PDF) set CTEQ6.6~\cite{cteq6.6} is used and the underlying event is modelled with \Jimmy~\cite{Jimmy,ATL-PHYS-PUB-2010-014}.  
The \WZ production cross section from $q\bar q'$ annihilation is calculated to be
$\WZtotXsecExpected^{+\WZtotXsecExpectedErrUp}_{\WZtotXsecExpectedErrDw}$ pb~\cite{Frixione:2002ik}.
Electroweak corrections have not been considered as they are not relevant at the current luminosity~\cite{Accomando:2001fn,Accomando:2005xp}.

Major backgrounds for \WZ signal detection are jets associated with $W$ or $Z$ gauge bosons, diboson and top-quark events.
\mcatnlo is used to model the $t\bar t$ and single top-quark events; to model the $W/Z$+jets and Drell-Yan backgrounds we use \alpgen~\cite{alpgen} for $e$ and $\mu$ decays and \pythia~\cite{pythia} for $\tau$ decays.
Events with heavy flavour dijets are modelled with \pythiaB~\cite{pythiab}.
The diboson processes $WW$ and $ZZ$ are modelled with \herwig and $W/Z+\gamma$ with \madgraph~\cite{madgraph}.
Whenever LO event generators are used, the cross sections are corrected 
by using k-factors to NLO or NNLO (if available) matrix element calculations~\cite{Frixione:2002ik}. 
The simulated background event samples generally correspond to an integrated luminosity of 1--10~\ifb.
Systematic uncertainties for simulated signal and backgrounds come from the uncertainty on their theoretical cross section, 
which are about 5-7\% for diboson processes and 8-9\% for $t\bar{t}$.

All event samples are simulated with multiple interactions in bunch crossings, including in-time and out-of-time pile-up, and with an average of
8 collision vertices per event. The number of interactions is re-weighted according to the luminosity distribution and the average number of interactions 
per bunch crossing of the data set used in this analysis, which varies according to the run period.

\section{Object Reconstruction}\label{sec:Reconstruction}
The main physics objects necessary to select \WZ events are electrons, muons, and \met.
To ensure that the lepton candidates originate from a primary vertex, the longitudinal impact parameter with respect to the primary vertex must be less than~{10 mm}. 
Combined muons are identified by matching tracks reconstructed in the muon spectrometer (MS) to tracks reconstructed in the inner detector (ID). Their momentum is calculated by combining information from the two tracks and correcting for the energy loss in the calorimeter. ID tracks that were tagged as a muon on the basis of matching with a segment in the MS are also included.
Only muons with $\pt > 15$ GeV and $|\eta| < 2.5$ are considered. The muon momentum in simulated events is smeared to account for a small difference between data and simulation. Non-prompt muons from hadronic jets are rejected by selecting only isolated muons, 
requiring the scalar sum of the track $\pt$ within $\Delta R <0.2$  of the muon\footnote{$\Delta R$ is defined as $\Delta R = \sqrt{(\Delta \eta)^2 + (\Delta \phi^2)}$.} to be less than 10\% of the muon $\pt$~\cite{ATLAS-CONF-2011-063}.

Electrons are formed by matching clusters found in the electromagnetic calorimeters to tracks in the inner detector.  Electron candidates must have $E_T > 15$ GeV, where $E_T$ is calculated from the cluster energy and track direction. To select central electrons and to avoid the transition regions between the calorimeters, the electron cluster must be within the regions $|\eta| < 1.37$ or $1.52 < |\eta| < 2.47$. 
Electrons are required to pass an electron identification requirement based on shower shape.
    To ensure isolation,  the sum of the calorimeter energy in a cone of $\Delta R = 0.3$ around the electron candidate, not including the energy of the candidate itself, must be less than 4 GeV. This quantity is corrected for the additional energy deposited in the presence of pile-up.
The electron energy in simulated events is smeared to account for a small difference between data and simulation. Electrons overlapping with selected muons within $\Delta R < 0.1$ are removed.

The \met $\mbox{}$ is calculated with reconstructed electrons within $|\eta| < 2.47$, muons within $|\eta |< 2.7$, and jets and calorimeter energy clusters outside of other reconstructed objects within $|\eta| < 4.5$.
The clusters are calibrated as electromagnetic or hadronic energy according to cluster topology.  A small correction avoids double-counting the energy deposited by muons in the calorimeters~\cite{ATLAS-CONF-2011-080}. 

\section{Event Selection}\label{sec:EventSelection}
At least one single electron or muon trigger is required to fire in order to
select the event, as described in Section~\ref{sec:ATLAS}. At least one primary vertex, with at least three good tracks associated, is required to remove non-collision backgrounds and ensure good object reconstruction. Events with two leptons of the same flavour and opposite charge with an invariant mass within 10 GeV of the $Z$ boson mass are selected. This reduces much of the background from multi-jet and top-quark production, and some diboson backgrounds.


Events are then required to have at least three reconstructed leptons
originating from the same primary vertex,
two leptons from a $Z\to\ell\ell$ decay and one additional third lepton.
The third lepton must pass more stringent identification criteria than the leptons attributed to the $Z$ boson and have $\pt >$ 20~GeV.
The additional identification criteria required for 
electrons are on the matched track quality, the ratio of the energy measured in the calorimeter to the momentum of the matched track, 
and the detection of transition radiation. Muons are required to be reconstructed as combined muons.

The transverse mass\footnote{The transverse mass $\MT$ is defined as $\MT^2 = 2E_{T\ell}E_{T\nu}-2{\bf p}_{T\ell}{\bf p}_{T\nu}$.} of the \W\ boson, $\MT^W$, is formed from the \met\ and the
third lepton. Furthermore,
the $\MT^W$ is required to be greater than 20 GeV. These cuts suppress
the remaining backgrounds from $Z$ and diboson production. 
For muon-triggered events, at least one of the muons is
required to have $\pt >$ 20 GeV and to have fired the muon trigger
to ensure that the trigger is well onto the efficiency plateau
above the threshold of the primary single muon trigger of 18 GeV. For electron-triggered events the $\pt$ requirement is 25~GeV and the primary single electron trigger has a threshold of 20 GeV.

\section{Signal Acceptance and Background Estimate}\label{sec:SignalAcceptance}
The acceptance of each cut is shown in Table~\ref{table:accWZMC} after all corrections have 
been applied to the simulated \WZ\ events decaying leptonically 
to electrons and muons in the final state. 

\begin{table}[htbp]
	 \centering
	 \caption{MC $\WZ\to\ell\nu\ell\ell$ signal acceptance after each cut, and the relative acceptance is shown in parentheses.}
	 \begin{tabular}{|l|c|c|c|c|}
	 \hline
	 	 Cut Sequence & \multicolumn{4}{|c|}{Acceptance [\%]} \\ \cline{2-5}
		\hline
	 	 Muon or electron trigger & \multicolumn{4}{|c|}{78.9 (78.9)} \\ \cline{2-5}
	 	 \hline
	 	 Primary Vertex & \multicolumn{4}{|c|}{78.7 (99.8)} \\ \cline{2-5}
		 \hline
	 	 $|m_{ll}-m_\Z| < 10\GeV$  & \multicolumn{4}{|c|}{ 28.2 (35.8)}  \\
	 	 \hline
	 	 Three leptons &  \multicolumn{4}{|c|}{12.3 (43.7)} \\
	 	 \hline
	 	 $\MET > 25\GeV$ & \multicolumn{4}{|c|}{10.0 (81.2)}\\
 	 	 \hline
	 	 $\MT^W > 20\GeV$  & \multicolumn{4}{|c|}{8.5 (84.9)}\\
	 	 \hline
	 	 Trigger Match &  \multicolumn{4}{|c|}{8.4 (99.5)}\\
	 	 \hline
	 \end{tabular}
          \label{table:accWZMC}
\end{table}

The dominant backgrounds are events with jets associated
with $Z$ bosons, diboson production, and top-quark events.  
Backgrounds are estimated from a data-driven method where possible
and from simulation otherwise.

The data driven method uses a sample of $Z$ events with an additional loose lepton 
which is dominated by $Z$+jet events from data. This sample contains two reconstructed
leptons passing all particle identification requirements and one
``lepton-like'' jet that fails to satisfy lepton quality (medium $e$) or
isolation ($\mu$) requirements. The two leptons which
pass all cuts must also pass the $Z$ reconstruction requirements. 
To select a control sample as close to the signal region as possible, 
all other event selection criteria, including the $\met$ and $\MT^W$
cuts are required.  The event yield is obtained by scaling each event 
in the resulting sample by the ``fake-factor'' $f(\pt)$, 
i.e. the probability that a ``lepton-like''
jet satisfies the quality or isolation requirements.  
The fake-factor is determined from a sample of events containing a $Z$ boson plus an extra
lepton-like jet, with a low missing energy requirement of $\met < 25\GeV$,
and extrapolated to high values of $\met$ using simulated
events. The validity of the extrapolation has been verified with
dijet events from simulation and data.

The background from $W/Z+\gamma$ events where the photon converts into an electron-positron pair
is not taken into account by the data-driven estimation methods, and is instead calculated 
with simulation. All other backgrounds are estimated using simulation.


For the electron and muon objects, the uncertainties associated with
the reconstruction and identification efficiency, energy scale, energy
smearing, and isolation are taken into account. The uncertainties are
determined from comparisons between simulated events and data in control
samples and are around 2\% to 6\% depending on the decay channel. 
The uncertainties on the objects that are used to calculate \met\ are used to calculate the systematic uncertainties on \met. Uncertainties in the description of the pile-up conditions by the simulation are also considered.
The total systematic uncertainty on the acceptance of the \met\ and transverse mass cuts due to the description by the simulation is 1--2\%.

\section{Results}\label{sec:Results}
The numbers of expected and observed events after applying all selection cuts
are shown in Table~\ref{ta:selected_data_MC}. Statistical uncertainties 
are given for all four trilepton channels. 
\WZNSigEventsObserved\, \WZ candidates are observed in data, with 
\WZNBkgEventsExpected$\pm$\WZNBkgEventsExpectedErrStat(stat)$^{+\WZNBkgEventsExpectedErrUpSys}_{-\WZNBkgEventsExpectedErrDwSys}$(sys)\ 
expected background events. The expected signal events include the contribution from $\tau$ lepton decays 
into electrons or muons, which accounts for \WZtau\ events.
The backgrounds from $\WW$ and multi-jet production were found to be negligible. 
Various kinematic variables and the $\W$ charge of \WZ candidates are shown in Figure~\ref{fig:wz-cand}.

\begin{table}[htbp]
  \centering
  \caption{Summary of observed events and expected signal and background
contributions in the four trilepton and combined channels.
 Statistical uncertainties are shown for the individual channels, and
both statistical and systematic uncertainties are shown for the
data-driven background estimation methods are
used for $W/Z+$jets for all decay channels. The last row shows the ratio of expected signal events over the background expectation.}
  \begin{tabular}{lccccc} 
    \hline\hline
     Final State    & $eee+\met$ & $ee\mu+\met$ & $e\mu\mu+\met$ &
     $\mu\mu\mu+\met$   & combined \\
    \midrule
    Observed        & \WZNSigEventsObservedEEE &
    \WZNSigEventsObservedEEM & \WZNSigEventsObservedEMM &
    \WZNSigEventsObservedMMM  & \WZNSigEventsObserved \\
    \midrule
    $ZZ$  
    & \WZNZZBkgEventsExpectedEEE$\pm$\WZNZZBkgEventsExpectedErrStatEEE
    & \WZNZZBkgEventsExpectedEEM$\pm$\WZNZZBkgEventsExpectedErrStatEEM   
    & \WZNZZBkgEventsExpectedEMM$\pm$\WZNZZBkgEventsExpectedErrStatEMM 
    &
    \WZNZZBkgEventsExpectedMMM$\pm$\WZNZZBkgEventsExpectedErrStatMMM
  & \WZNZZBkgEventsExpected$\pm$\WZNZZBkgEventsExpectedErrStat$\pm$\WZNZZBkgEventsExpectedErrSys \\
    \vspace{2 mm} $W/Z+$jets     
    & \WZNZjetsDDBkgEventsExpectedEEE$\pm$\WZNZjetsDDBkgEventsExpectedErrStatEEE
    & \WZNZjetsDDBkgEventsExpectedEEM$\pm$\WZNZjetsDDBkgEventsExpectedErrStatEEM   
    & \WZNZjetsDDBkgEventsExpectedEMM$\pm$\WZNZjetsDDBkgEventsExpectedErrStatEMM 
    &
    \WZNZjetsDDBkgEventsExpectedMMM$\pm$\WZNZjetsDDBkgEventsExpectedErrStatMMM
 & \WZNZjetsDDBkgEventsExpected$\pm$\WZNZjetsDDBkgEventsExpectedErrStat$^{+\WZNZjetsDDBkgEventsExpectedErrSysUp}_{-\WZNZjetsDDBkgEventsExpectedErrSysDown}$ \\
    Top           
    & \WZNTopBkgEventsExpectedEEE$\pm$\WZNTopBkgEventsExpectedErrStatEEE
    & \WZNTopBkgEventsExpectedEEM$\pm$\WZNTopBkgEventsExpectedErrStatEEM   
    & \WZNTopBkgEventsExpectedEMM$\pm$\WZNTopBkgEventsExpectedErrStatEMM 
    &
    \WZNTopBkgEventsExpectedMMM$\pm$\WZNTopBkgEventsExpectedErrStatMMM
 & \WZNTopBkgEventsExpected$\pm$\WZNTopBkgEventsExpectedErrStat$\pm$\WZNTopBkgEventsExpectedErrSys \\   
    $W/Z+\gamma$  
    & \WZNZgammaBkgEventsExpectedEEE$\pm$\WZNZgammaBkgEventsExpectedErrStatEEE
    & --   
    & \WZNZgammaBkgEventsExpectedEMM$\pm$\WZNZgammaBkgEventsExpectedErrStatEMM
    & -- 
    & \WZNZgammaBkgEventsExpected$\pm$\WZNZgammaBkgEventsExpectedErrStat$\pm$\WZNZgammaBkgEventsExpectedErrSys \\ 
    \midrule
    Total Background        & \WZNBkgEventsExpectedEEE$\pm$\WZNBkgEventsExpectedErrStatEEE
    & \WZNBkgEventsExpectedEEM$\pm$\WZNBkgEventsExpectedErrStatEEM
    & \WZNBkgEventsExpectedEMM$\pm$\WZNBkgEventsExpectedErrStatEMM
    &
    \WZNBkgEventsExpectedMMM$\pm$\WZNBkgEventsExpectedErrStatMMM
    & \WZNBkgEventsExpected$\pm$\WZNBkgEventsExpectedErrStat$^{+\WZNBkgEventsExpectedErrUpSys}_{-\WZNBkgEventsExpectedErrDwSys}$ \\
    \midrule
    Expected Signal        
    & \WZNSigEventsExpectedEEE$\pm$\WZNSigEventsExpectedErrStatEEE    
    & \WZNSigEventsExpectedEEM$\pm$\WZNSigEventsExpectedErrStatEEM 
    & \WZNSigEventsExpectedEMM$\pm$\WZNSigEventsExpectedErrStatEMM 
    &
    \WZNSigEventsExpectedMMM$\pm$\WZNSigEventsExpectedErrStatMMM
    & \WZNSigEventsExpected$\pm$\WZNSigEventsExpectedErrStat$\pm$\WZNSigEventsExpectedErrSys \\
    \hline
   \midrule
   Expected S/B             & \WZSBEEE   & \WZSBEEM  & \WZSBEMM  & \WZSBMMM    & \WZSB \\ 
    \hline\hline
  \end{tabular}
  \label{ta:selected_data_MC}
\end{table}

 \begin{figure}[htbp]
 \begin{center}
 \subfigure[]{
 \includegraphics[width=0.33\textwidth]{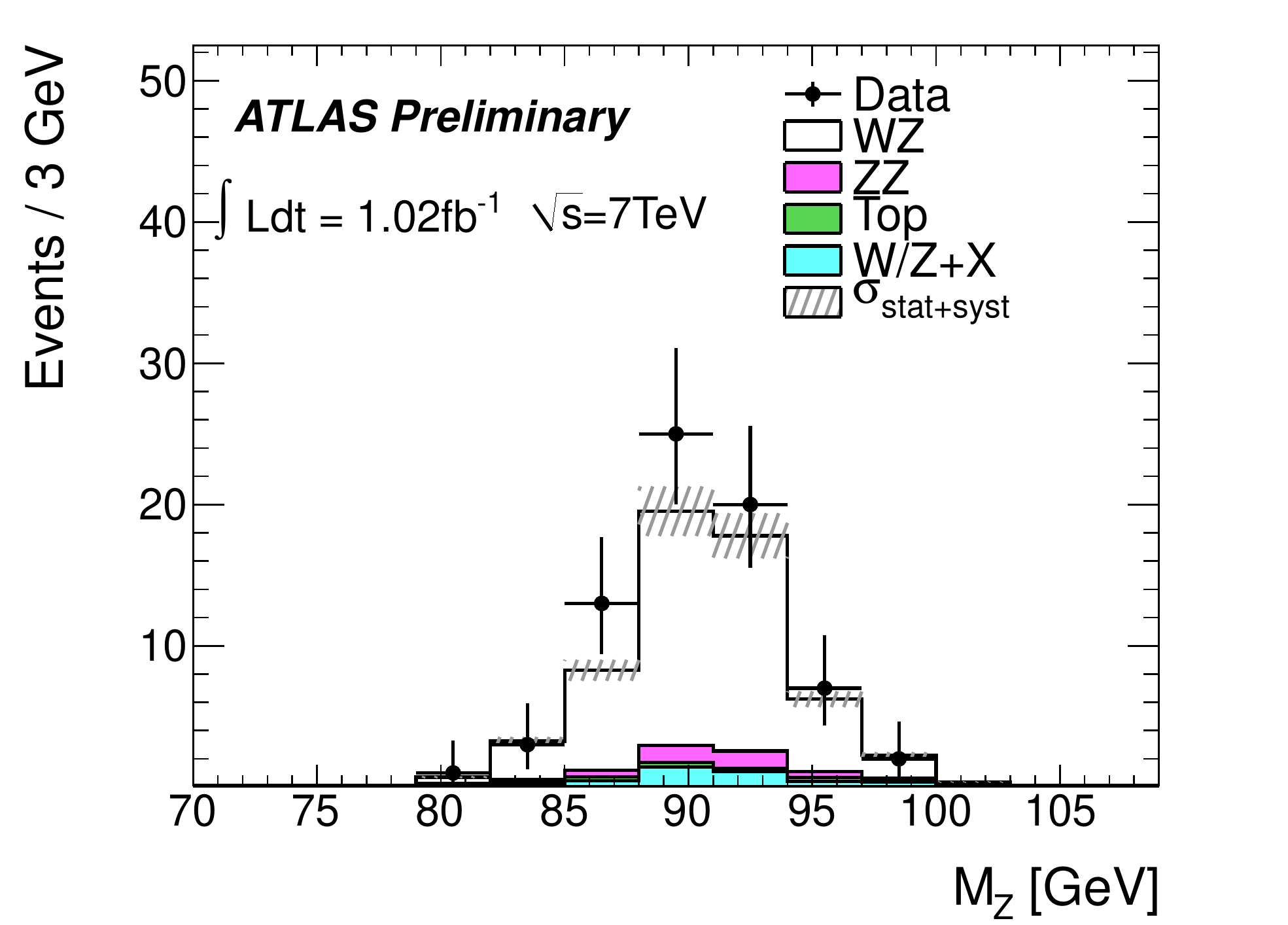}
 }\hskip -.4cm
 \subfigure[]{
 \includegraphics[width=0.33\textwidth]{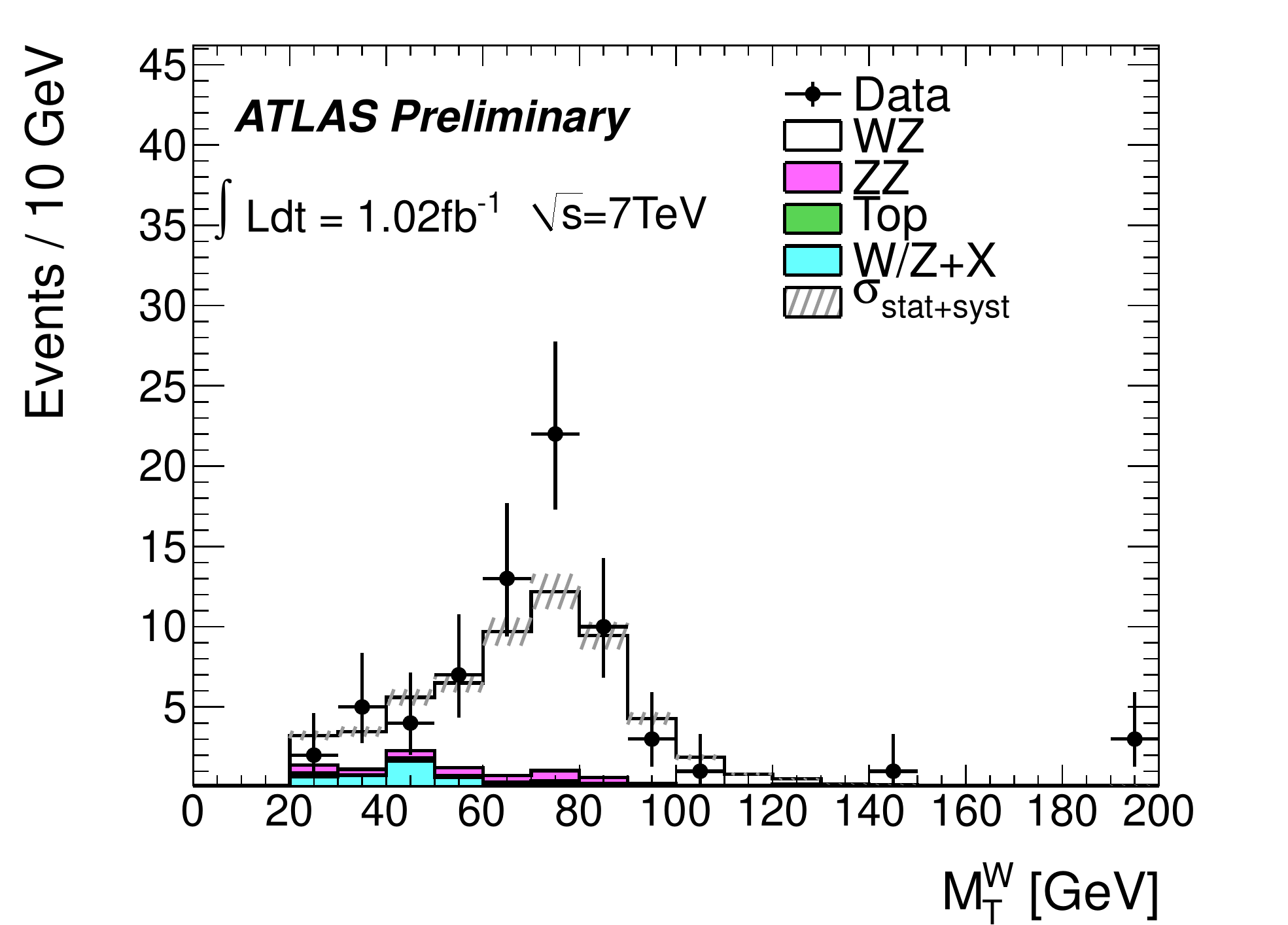}\hfill \hskip -.4cm
 }
 \subfigure[]{
 \includegraphics[width=0.33\textwidth]{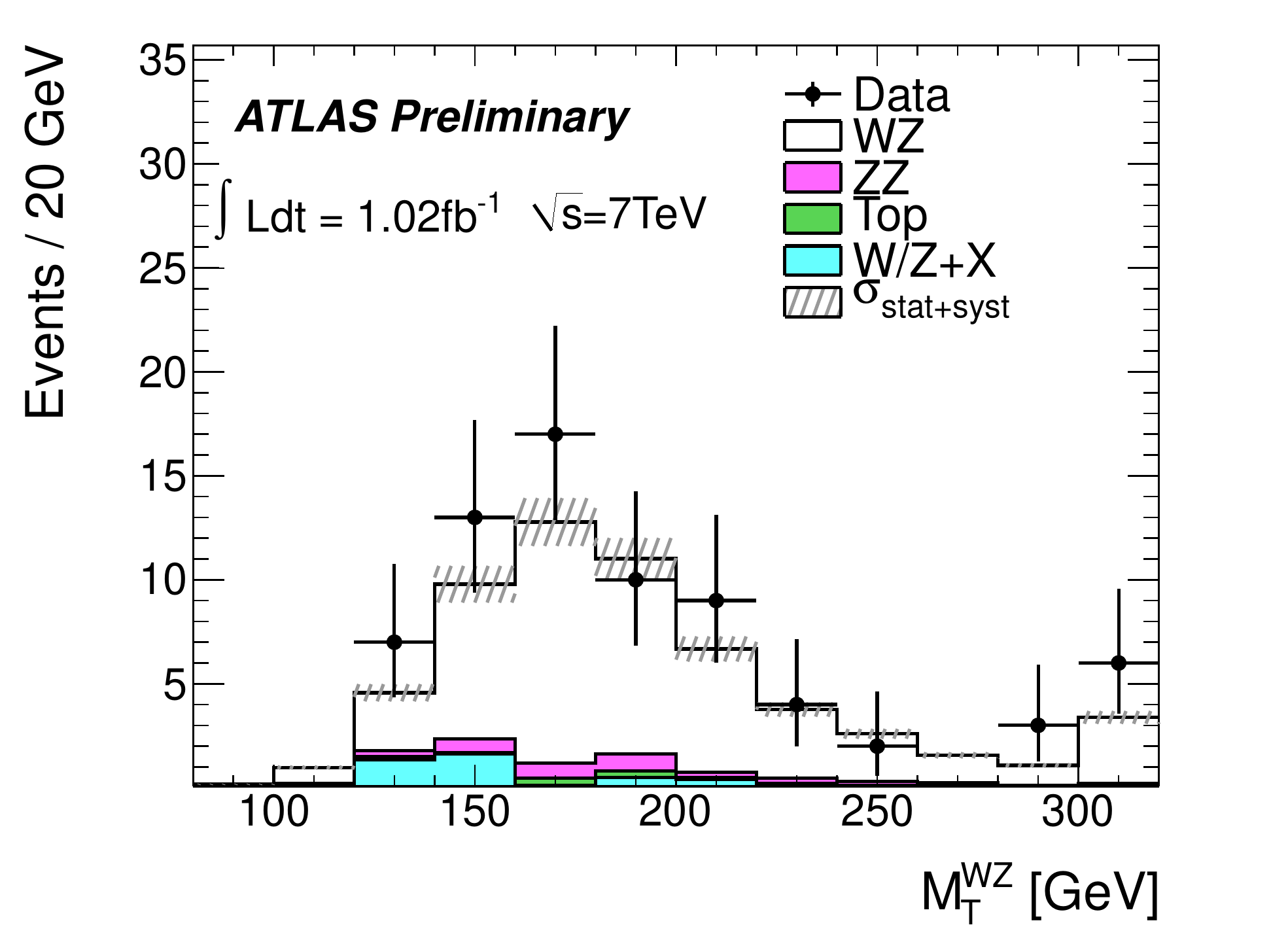}\hfill 
 }
 \subfigure[]{
 \includegraphics[width=0.33\textwidth]{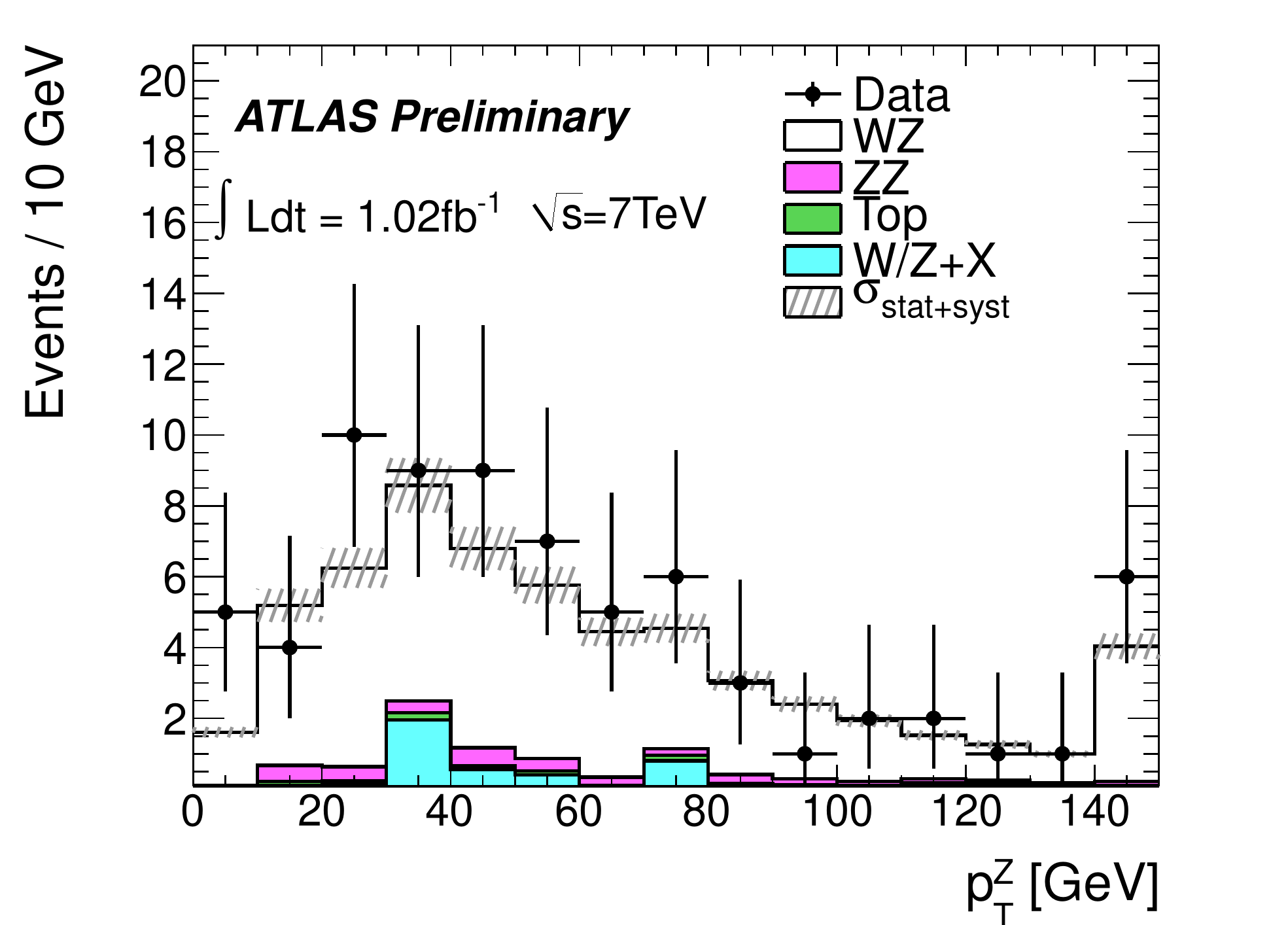}
 }\hskip -.4cm
 \subfigure[]{
 \includegraphics[width=0.33\textwidth]{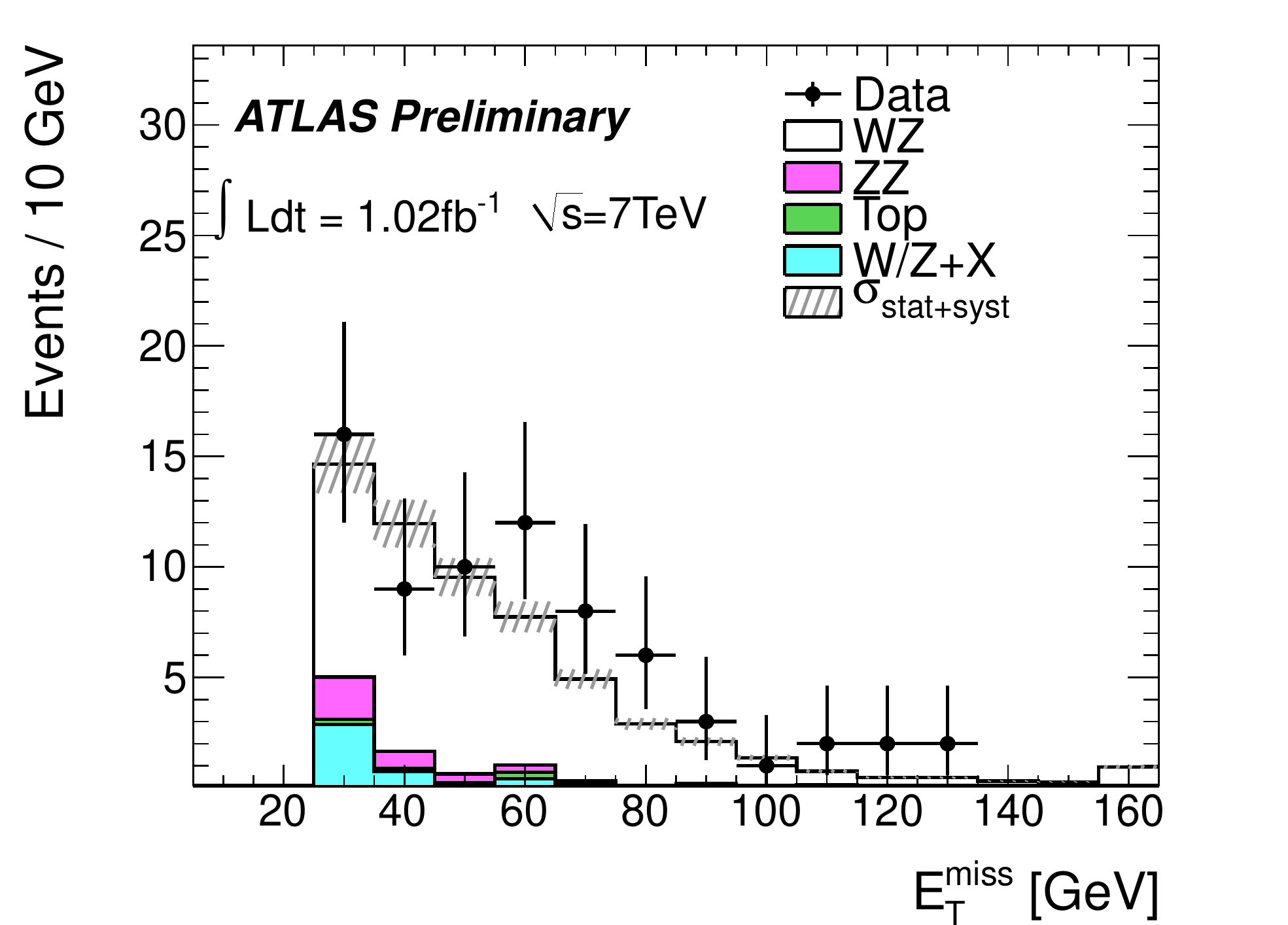}\hfill \hskip -.4cm 
 }
 \subfigure[]{
 \includegraphics[width=0.33\textwidth]{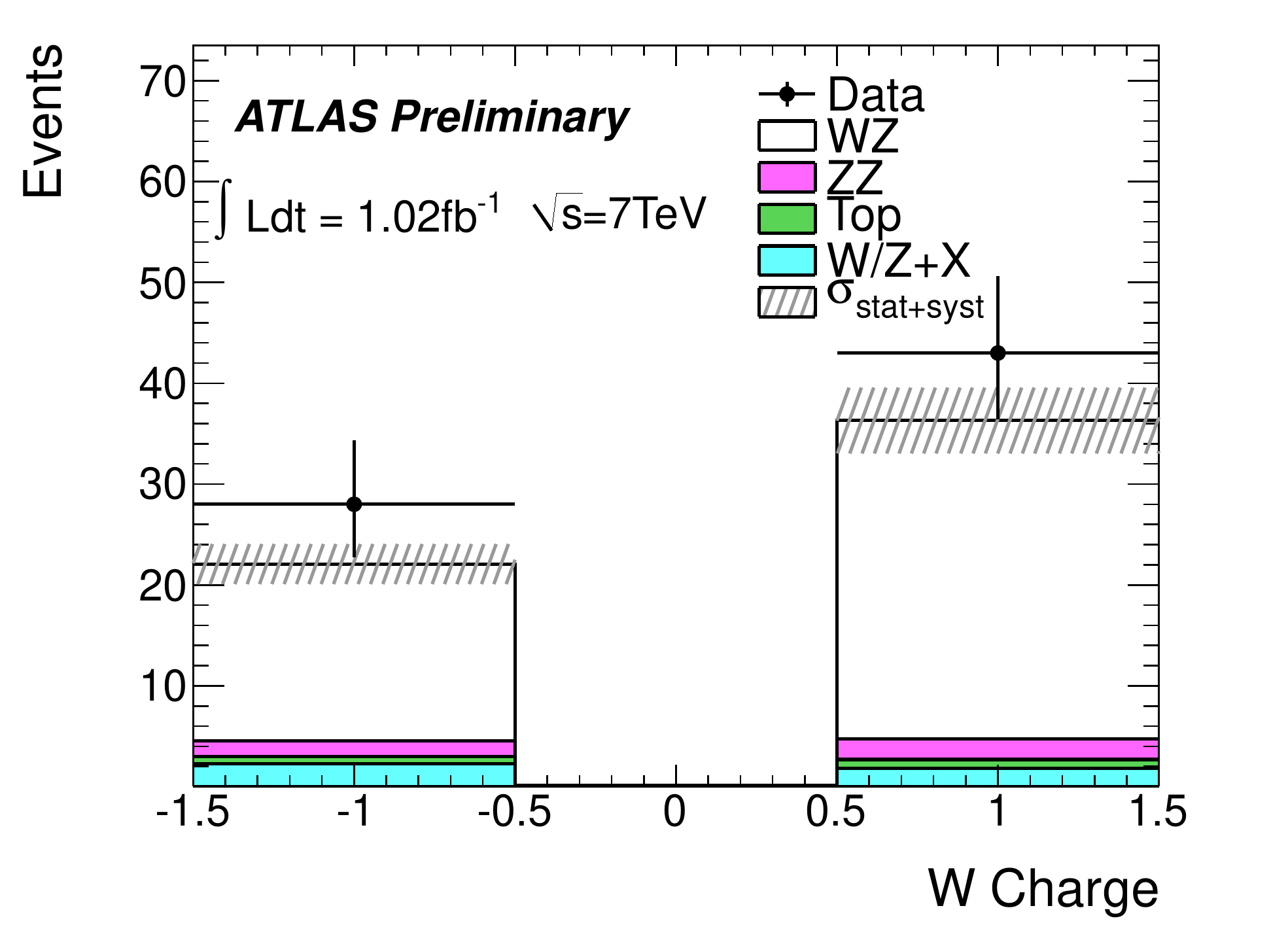}\hfill
 }
 \caption{\small Distributions of various kinematic variables, number of leptons, jets, and the \W charge of \WZ candidates. The kinematic variables shown are: 
     (a) invariant mass of $Z$,
     (b) transverse mass of \W,
     (c) transverse mass of \WZ,
     (d) transverse momentum of the $Z$ ($\pt^{Z}$), 
     (e) \met \mbox{},
     (f) \W charge, 
    The points represent observed event counts with statistical errors, 
whereas the stacked histograms are the predictions from simulation including the statistical and systematic uncertainty. The last bin is an overflow bin.
   }
 \label{fig:wz-cand}
 \end{center}
 \end{figure} 

\subsection{Cross Section Extraction} \label{subsec:CrossSection}
In order to combine the different channels, a common phase space region is defined in which a fiducial cross section is extracted. The common phase space is defined as $\pt^{\mu,e} > 15\GeV$, $|\eta^{\mu,e}| < 2.5$, $\pt^{\nu} > 25\GeV$, $|M_{\ell\ell}-M_{\rm Z}| < 10\GeV$, and 
$M_{\mathrm{T}}^{W} > 20\GeV$, to approximate the event selection used in this analysis.

For a given channel $\WZ\to\ell\nu\ell\ell$ where $\ell \in \{e,\mu\}$, we define the fiducial cross section to be
\begin{equation}\label{eq:xsecfid}
\sigma_{WZ}^{fid}\times \mathcal{B}(WZ\to\ell\nu\ell\ell ) = \sigma_{WZ\to\ell\nu\ell\ell}^{fid} = \frac{ N^{obs}_{\ell\nu\ell\ell} - N^{bkg}_{\ell\nu\ell\ell}}{\mathcal{L} \times C_{WZ\to\ell\nu\ell\ell}}\times(1-\frac{N^{MC}_{\tau}}{N^{MC}_{sig}})
\end{equation}
Here, $N^{obs}_{\ell\nu\ell\ell}$ and $N^{bkg}_{ \ell\nu\ell\ell}$ denote the
number of observed and background events respectively, $\mathcal{L}$
is the integrated luminosity and $C_{WZ\to\ell\nu\ell\ell}$ is the efficiency corrected to account for measured differences in trigger and reconstruction
efficiencies between simulated and data samples and for the
extrapolation to the fiducial volume. 
$\mathcal{B}(WZ\to\ell\nu\ell\ell)$ is the branching ratio for a \W~ to decay to
$\ell\nu$ and a $Z$ to decay to $\ell\ell$, $N^{MC}_{\tau}$ is the number of \WZ\ events where at least one of the bosons decays tauonically and
$N^{MC}_{\mathrm{sig}}$ is the number of \WZ\ events decaying into any lepton flavour.
Since the fiducial volume is
defined by the leptonic kinematics, the calculated cross section must include the branching ratio. 
The contribution from $\tau$ lepton decays, about $4\%$, is not included and explicitly removed in the fiducial cross section definition
with a scaling factor derived from the fraction of expected events from $\tau$ decays.

In addition to the fiducial cross section, we also calculate a
cross section in the total phase space in each channel. The
total cross section is calculated as: 
\begin{equation}\label{eq:xsectot}
\sigma_{WZ}^{tot} = \frac{ (N^{obs}_{\ell\nu\ell\ell} - N^{bkg}_{\ell\nu\ell\ell})}{\mathcal{L} \times \mathcal{B}(WZ\to\ell\nu\ell\ell ) \times A_{WZ\to\ell\nu\ell\ell} \times C_{WZ\to\ell\nu\ell\ell}}\times(1-\frac{N^{MC}_{\tau}}{N^{MC}_{sig}})
\end{equation}
where $A_{WZ\to\ell\nu\ell\ell}$ is the ratio of the number of events
within the fiducial phase space region to the total number of
generated events, needed to extrapolate to the total cross section for
each channel, calculated using the event generator information.  
This ratio $A_{WZ\to\ell\nu\ell\ell}$ equals 
0.389, 0.392, 0.392, and 0.389 
for the $\mu\mu\mu$, $e\mu\mu$, $ee\mu$, and $eee$ channels, respectively.
The uncertainties on $A_{WZ\to\ell\nu\ell\ell}$ due to the MC signal sample statistics are
estimated to be about 2\% (uncorrelated among channels) and are included in the systematic uncertainties. 
The uncertainty due to the parton distribution functions is calculated by combining the uncertainty evaluated from
44 CTEQ6.6 error eigenvectors (1.5\%), and the central value deviation from 
MSTW2008 NLO (0.6\%). The combined systematic uncertainty
with quadratic sum is 1.6\%.

In the actual calculation, loglikelihood method is used to fit the cross-section and add nuisance parameters to account 
for the systematic uncertainty on the number of expected signal and background events in each channel, and to find the most probable 
value of $\sigma$ (fiducial or total) the negative log-likelihood function is simultaneously minimized over $\sigma$ and all the nuisance parameters  $x_k$.

The final results for the combined fiducial cross section for
the \WZ bosons decaying directly into electrons and muons (excluding contributions
from $\tau$ lepton decays) and the combined total inclusive cross section measurements are shown below.  
The fractional signal contribution from tau decays has been estimated with the simulated
signal sample and is removed from the fiducial cross section

\begin{center}
$\sigma_{WZ\to\ell\nu\ell\ell}^{fid} = \WZfidXsec^{+\WZfidXsecStatErrUp}_{-\WZfidXsecStatErrDw}$(stat) $^{+\WZfidXsecSysErrUp}_{-\WZfidXsecSysErrDw}$(syst) $^{+\WZfidXsecLumiErrUp}_{-\WZfidXsecLumiErrDw}$(lumi)~fb

\vspace{0.4cm}
$\sigma_{WZ}^{tot} =  \WZtotXsec^{+\WZtotXsecStatErrUp}_{-\WZtotXsecStatErrDw}$(stat) $^{+\WZtotXsecSysErrUp}_{-\WZtotXsecSysErrDw}$(syst) $^{+\WZtotXsecLumiErrUp}_{-\WZtotXsecLumiErrDw}$(lumi)~\pb.
\end{center}

\subsection{Limits on Anomalous Triple Gauge Couplings} \label{subsec:TGC}
The model-independent method of effective action and effective Lagrangians allows for less restricted theories compared to the SM to be contructed. Expressions for the most general effective Lagrangian for a TGC vertex with two charged and one neutral vector boson can be found in Refs. \cite{Hagiwara} or \cite{EllisonWudka}. If only terms that separately conserve charge conjugation $C$ and parity $P$ are considered then the Lagrangian reduces to
\begin{equation}
\label{eqn:Lagrangian}
\frac{\mathcal{L}_{WWZ}}{g_{WWZ}}=i\left[g^{Z}_{1}(W^{\dagger}_{\mu\nu}W^{\mu}Z^{\nu}-W_{\mu\nu}W^{\dagger\mu}Z^{\nu})+\kappa^{Z}W^{\dagger}_{\mu}W_{\nu}Z^{\mu\nu}+\frac{\lambda}{m^{2}_{W}}W^{\dagger}_{\rho\mu}W^{\mu}_{\nu}Z^{\nu\rho}\right]
\end{equation} 
where $g_{WWZ}=-e\cot\theta_{W}$, $\theta_{W}$ is the weak mixing angle, $X_{\mu\nu}=\partial_{\mu}X_{\nu}-\partial_{\nu}X_{\mu}$ and $g^{Z}_{1}$,  $\kappa^{Z}$ and $\lambda$ are the coupling parameters that will be constrained in this analysis. It should be noted that $g^{Z}_{1}$, $\kappa^{Z}$ and $\lambda$ are dimensionless coupling parameters that multiply dimension-four and dimension-six operators respectively; higher dimension operators are neglected. To recover the SM result $g_{1}^{Z}=1$, $\kappa^{Z}=1$, $\lambda=0$. In \WZ production $g^{Z}_{1}$ and $\lambda$ are proportional to $\hat{s}$, whereas $\kappa^{Z}$ is proportional to $\sqrt{\hat{s}}$. The cross section has a quadratic dependance on a given coupling.

To avoid tree-level unitarity violation at high centre-of-mass energies, which arises when radiative corrections from this new effective Lagrangian are bigger than the tree-level contributions, the anomalous couplings must vanish as $\sqrt{\hat{s}}\rightarrow\infty$. To achieve this an arbitrary form factor is introduced according to
\begin{equation}
\alpha(\hat{s})=\frac{\alpha_{0}}{(1+\hat{s}/\Lambda^{2})^{2}}
\end{equation}
where $\hat{s}$ is the invariant mass of the vector boson pair, $\alpha$ stands for $\Delta g^{Z}_{1}$, $\Delta \kappa^{Z}$ or $\lambda$, the deviations of the anomalous couplings from the Standard Model values, and $\alpha_{0}$ is the value of the anomalous coupling at low energy. A dipole form factor was used and $\Lambda$, the `scale of new physics', was chosen to be 3 TeV.

\subsubsection{Reweighting and Limit Setting Procedure}

In MC@NLO version 4.0 \cite{mc@nlo4manual} it is possible to generate \WZ events at any non-SM phase space point. Each event gets a vector of ten weights, $w_0-w_9$, associated to it which allow an event sample to be reweighted to another phase space point. The weight at this new point is given by

\begin{eqnarray} 
\label{eqn:weight}
w_{\mathrm{TOT}} &\propto& w_0 + 2\Delta g^Z_1w_1 + 2\Delta\kappa^Zw_2 + 2\lambda w_3 \nonumber\\
&& + 2\Delta g^Z_1\Delta\kappa^Zw_4 + 2\Delta g^Z_1\lambda w_5 + 2\Delta\kappa^Z\lambda w_6 \nonumber\\
&& + (\Delta g^Z_1)^2w_7 + (\Delta\kappa^Z)^2w_8 + (\lambda)^2 w_9.
\end{eqnarray}

To set limits on aTGCs a profile likelihood test with a frequentist limit approach ~\cite{frequentist} was adopted to determine all possible cross sections consistent with the observed data at the 95\% confidence level (C.L.), giving upper and lower confidence limits for the cross section. The reweighting procedure allows the cross section to be determined as a function of aTGCs. Values of the aTGCs which predict cross sections inside of the 95\% confidence interval (C.I.) of the cross section determine the 95\% C.I. of the anomalous couplings. The C.I. for each anomalous coupling was determined separately with the other couplings set to their SM value. It should be noted that fiducial cross section values were used in this limit setting procedure, hence changes in $A_{WZ\rightarrow\ell\nu\ell\ell}$ are already taken into account. The effect of changes in $C_{WZ\rightarrow\ell\nu\ell\ell}$ within the range of couplings is $\sim$1\% and was not taken into account.

\subsubsection{Expected and Observed Limits}

Assuming SM values, the expected experimental limits on aTGCs can be found in Table \ref{table:limExp}, where the median anomalous coupling C.I. is the best estimate, and the errors represent the 68\% C.L. band of the anomalous coupling limits. The observed limits at the 68\% and  the 95\% C.I on the anomalous couplings $\Delta g_1^Z$, $\Delta \kappa^Z$, and $\lambda$ can be found in Table \ref{table:limObs}. For 68\% C.I. two disjunct intervals are found. The observed 95\% C.I. limits 
include the SM expectations and a comparison of ATLAS and Tevatron results are shown in Figure~\ref{fig:combined}; LEP results from $W^+W^-$ production can be found in Ref.~\cite{LEPlimits}. Significant improvement in these limits is expected with more integrated luminosity and refined extraction methods taking advantage of the differential spectrum of kinematic quantities. 

The influence of the scale $\Lambda$ on the observed limits was estimated by setting $\Lambda$ above the centre-of-mass energy and calculating the expected cross section at 
the observed limits. The increase of the cross section is between 2\% and 7\% depending on the energy dependence of the coupling parameters. This will result in tighter
limits on the coupling values. The size of this effect is estimated to be about 10\% on $\lambda$, and less on $\Delta\kappa$ and $g^Z_1$.  

\begin{table}[htbp]
\renewcommand{\arraystretch}{1.4}
	 \centering
	 \caption{Expected experimental limits at the 95\% C.I. on the anomalous couplings $\Delta g_1^Z$, $\Delta \kappa^Z$, and $\lambda$, assuming SM values. The error bars indicate the 68\% C.L. of the expected limits including systematic uncertainties for each coupling at the 95\% C.I..}
	 \begin{tabular}{|l|c|}
	 \hline
	 Anomalous Coupling & Limits of the 95\% C.I. \\
	 \hline
	 \hline

	 $\Delta g_1^Z$ & $\left[\expglow^{+\errupexpglow}_{-\errlowexpglow}, \expgup^{-\errlowexpgup}_{+\errupexpgup} \right]$\\

	 \hline

	 $\Delta \kappa^Z$ & $\left[\expkappalow^{+\errupexpkappalow}_{-\errlowexpkappalow}, \expkappaup^{-\errlowexpkappaup}_{+\errupexpkappaup} \right]$\\

	 \hline

	 $\lambda$ & $\left[\explambdalow^{+\errupexplambdalow}_{-\errlowexplambdalow}, \explambdaup^{-\errlowexplambdaup}_{+\errupexplambdaup} \right]$\\

	 \hline
	 \end{tabular}
          \label{table:limExp}
\end{table}

\begin{table}[htbp]
\renewcommand{\arraystretch}{1.4}
	 \centering
	 \caption{Observed limits at the 68\% and 95\% C.I. on the anomalous couplings $\Delta g_1^Z$, $\Delta \kappa^Z$, and $\lambda$.}
	 \begin{tabular}{|l|c|c|}
	 \hline
	 Anomalous Coupling & Limits of the 68\% C.I. & Limits of the 95\% C.I. \\
	 \hline
	 \hline

	 $\Delta g_1^Z$ &   $\left[\obsglowSixtyEightA,  \obsgupSixtyEightA\right],  \left[\obsglowSixtyEightB,  \obsgupSixtyEightB\right]$ & $\left[\obsglow,  \obsgup\right]$\\

	 \hline

	 $\Delta \kappa^Z$ & $\left[\obskappalowSixtyEightA,  \obskappaupSixtyEightA\right], \left[\obskappalowSixtyEightB,  \obskappaupSixtyEightB\right]$ & $\left[\obskappalow,  \obskappaup\right]$\\

	 \hline

	 $\lambda$ &  $ \left[\obslambdalowSixtyEightA,  \obslambdaupSixtyEightA\right], \left[\obslambdalowSixtyEightB,  \obslambdaupSixtyEightB\right]$ & $\left[\obslambdalow,  \obslambdaup\right]$\\

	 \hline
	 \end{tabular}
          \label{table:limObs}
\end{table}

 \begin{figure}[htbp]
 \begin{center}
 \includegraphics[width=.6\textwidth]{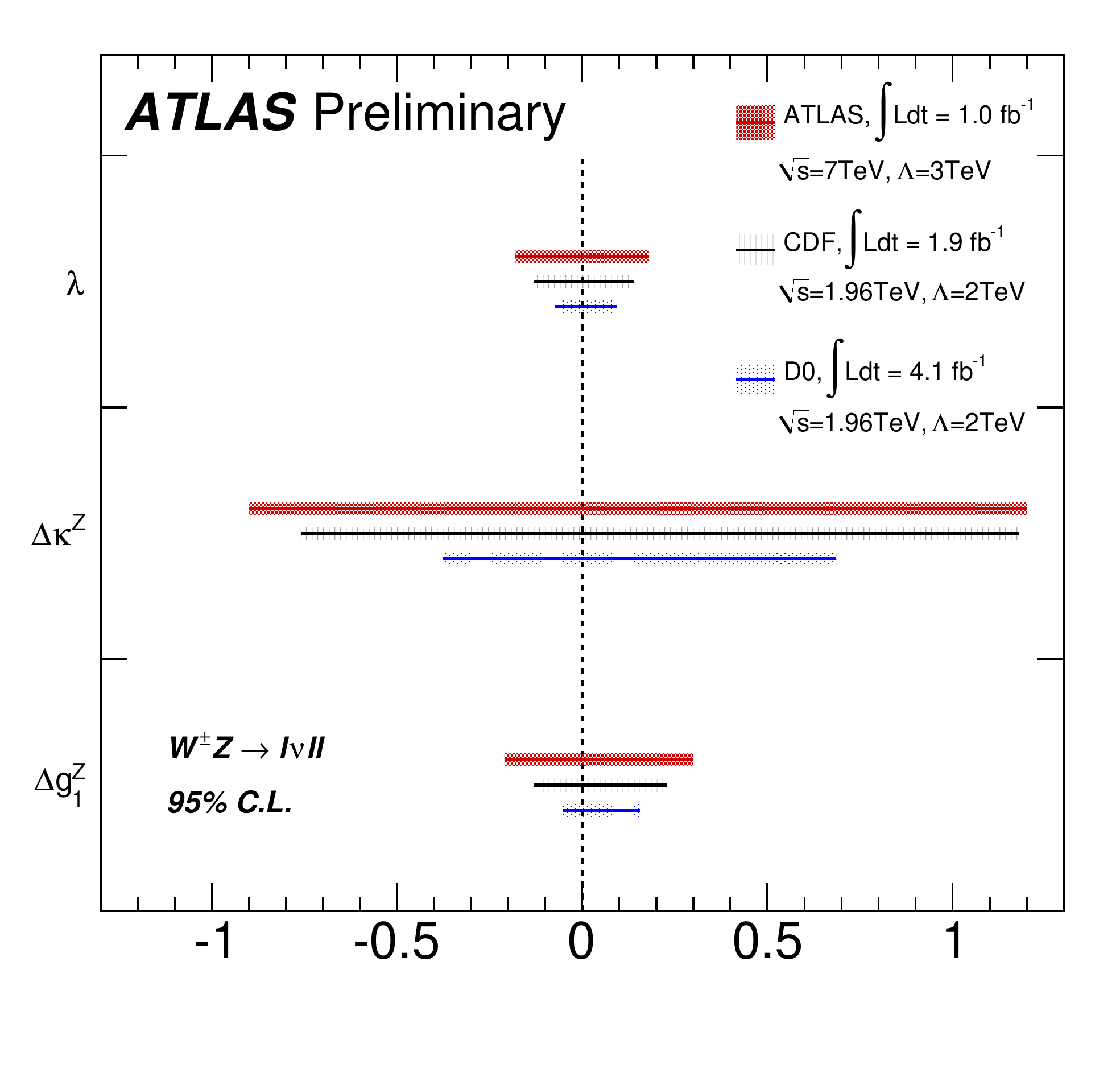}
 \caption{\small Shown are aTGC limits from ATLAS and Tevatron experiments. CDF~\cite{CDFlimits} and D\O~\cite{D0limits} limits are for \WZ production with a $p_T(Z)$ shape fit; ATLAS limits are for a cross section fit. Luminosities, centre-of-mass energy and cut-off $\Lambda$ for each experiment are shown and the limits are for 95\% C.I..}
 \label{fig:combined}
 \end{center}
 \end{figure}

\section{Conclusion} \label{subsec:Conclusion}
A measurement of the \WZ production cross section with the ATLAS
detector in LHC proton-proton collisions at $\sqrt{s}=7$ TeV has
been performed using final states with electrons and muons.
With an integrated luminosity of \lumi~\ifb\ a total of \WZNSigEventsObserved\ candidates is observed with a background
expectation of \WZNBkgEventsExpected$\pm$\WZNBkgEventsExpectedErrStat(stat)$^{+\WZNBkgEventsExpectedErrUpSys}_{-\WZNBkgEventsExpectedErrDwSys}$(sys). 
The Standard Model expectation for the number of signal events is \WZNSigEventsExpected$\pm$\WZNSigEventsExpectedErrStat(stat)$\pm$\WZNSigEventsExpectedErrSys(sys). 
The fiducial and total cross sections were determined to be \\
$\sigma_{WZ\to \ell\nu \ell\ell}^{fid} = \WZfidXsec^{+\WZfidXsecStatErrUp}_{-\WZfidXsecStatErrDw}$(stat) $^{+\WZfidXsecSysErrUp}_{-\WZfidXsecSysErrDw}$(syst) $^{+\WZfidXsecLumiErrUp}_{-\WZfidXsecLumiErrDw}$(lumi)~fb\ 
and\ 
$\sigma_{WZ}^{tot} =  \WZtotXsec^{+\WZtotXsecStatErrUp}_{-\WZtotXsecStatErrDw}$(stat) $^{+\WZtotXsecSysErrUp}_{-\WZtotXsecSysErrDw}$(syst) $^{+\WZtotXsecLumiErrUp}_{-\WZtotXsecLumiErrDw}$(lumi)~\pb, respectively.
The total cross section is in good agreement with the SM expectation~\cite{Frixione:2002ik} of $\WZtotXsecExpected^{+\WZtotXsecExpectedErrUp}_{\WZtotXsecExpectedErrDw}$ pb. 
Limits on the anomalous triple gauge couplings $g^{Z}_{1}$, $\kappa^Z$ and $\lambda$ are derived.



\bigskip 
\bibliography{atlas_wz}

\providecommand{\href}[2]{#2}\begingroup\raggedright\begin{thebibliography}{10}

\bibitem{DKS:1999}
L.~Dixon, Z.~Kunzst, and A.~Signer, {\em {Vector Boson Production in Hadron
  Collisions at order ${\alpha}_{s}$: Lepton Correlations and Anomalous
  Couplings}\/},  Phys. Rev. D {\bf 60} (1999)  114037,
  \href{http://arxiv.org/abs/hep-ph/9907305v1}{{\tt arXiv:hep-ph/9907305v1}}.

\bibitem{PhysRevD.65.094041}
K.~L. Adamson, D.~de~Florian, and A.~Signer, {\em Gluon induced contributions
  to $WZ$ and $W\gamma{}$ production at NNLO\/},
  \href{http://dx.doi.org/10.1103/PhysRevD.65.094041}{Phys. Rev. D {\bf 65}
  (2002)  094041}.

\bibitem{atlas}
{ATLAS Collaboration}, {\em {The ATLAS experiment at the CERN Large Hadron
  Collider}\/},  {JINST} {\bf 3 S08003} (2008)  .

\bibitem{Frixione:2002ik}
S.~Frixione and B.~R. Webber, {\em {Matching NLO QCD computations and parton
  shower simulations}\/},  JHEP {\bf 06} (2002)  029,
\href{http://arxiv.org/abs/arXiv:hep-ph/0204244}{{\tt arXiv:hep-ph/0204244}}.

\bibitem{Herwig}
G.~Corcella et al., {\em {HERWIG 6: an event generator for Hadron Emission
  Reactions With Interfering Gluons (including supersymmetric processes)}\/},
  JHEP {\bf 01} (2001)  010,
\href{http://arxiv.org/abs/hep-ph/0011363}{{\tt arXiv:hep-ph/0011363}}.

\bibitem{cteq6.6}
P.~M. Nadolsky et al., {\em {Implications of CTEQ global analysis for collider
  observables}\/},  \href{http://dx.doi.org/10.1103/PhysRevD.78.013004}{Phys.
  Rev. D {\bf 78} (2008)  013004},
\href{http://arxiv.org/abs/hep-ph/0802.0007}{{\tt arXiv:hep-ph/0802.0007}}.

\bibitem{Jimmy}
J.~M. Butterworth, J.~R. Forshaw, and M.~H. Seymour, {\em {Multiparton
  interactions in photoproduction at HERA}\/},
  \href{http://dx.doi.org/10.1007/s002880050286}{Z. Phys. {\bf C72} (1996)
  637},
\href{http://arxiv.org/abs/hep-ph/9601371}{{\tt arXiv:hep-ph/9601371}}.

\bibitem{ATL-PHYS-PUB-2010-014}
T.~A. Collaboration, {\em First tuning of HERWIG/JIMMY to ATLAS data\/},
  {ATLAS Note} {\bf ATL-PHYS-PUB-2010-014} (2010)  .

\bibitem{Accomando:2001fn}
E.~Accomando, A.~Denner, and S.~Pozzorini, {\em {Electroweak correction effects
  in gauge boson pair production at the CERN LHC}\/},
  \href{http://dx.doi.org/10.1103/PhysRevD.65.073003}{Phys. Rev. {\bf D65}
  (2002)  073003},
\href{http://arxiv.org/abs/hep-ph/0110114}{{\tt arXiv:hep-ph/0110114}}.

\bibitem{Accomando:2005xp}
E.~Accomando and A.~Kaiser, {\em {Electroweak corrections and anomalous triple
  gauge-boson couplings in W+ W- and W+- Z production at the LHC}\/},
  \href{http://dx.doi.org/10.1103/PhysRevD.73.093006}{Phys. Rev. {\bf D73}
  (2006)  093006},
\href{http://arxiv.org/abs/hep-ph/0511088}{{\tt arXiv:hep-ph/0511088}}.

\bibitem{alpgen}
M.~L. Mangano, M.~Moretti, F.~Piccinini, R.~Pittau, and A.~Polosa, {\em ALPGEN,
  a generator for hard multiparton processes in hadronic collisions\/},  JHEP
  {\bf 07} (2003)  001, \href{http://arxiv.org/abs/hep-ph/0206293}{{\tt
  arXiv:hep-ph/0206293}}.

\bibitem{pythia}
T.~Sjostrand et al., {\em {High-energy physics event generation with PYTHIA
  6.1}\/},  \href{http://dx.doi.org/10.1016/S0010-4655(00)00236-8}{Comput.
  Phys. Commun. {\bf 135} (2001)  238},
\href{http://arxiv.org/abs/hep-ph/0010017}{{\tt arXiv:hep-ph/0010017}}.

\bibitem{pythiab}
S.~P. Baranov and M.~Smizanska, {\em {Semihard b quark production at
  high-energies versus data and other approaches}\/},
\href{http://dx.doi.org/10.1103/PhysRevD.62.014012}{Phys. Rev. D {\bf 62}
  (2000)  014012}.

\bibitem{madgraph}
J.~Alwall et al., {\em {MadGraph/MadEvent v4: The New Web Generation}\/},  JHEP
  {\bf 09} (2007)  028,
\href{http://arxiv.org/abs/hep-ph/0706.2334}{{\tt arXiv:hep-ph/0706.2334}}.

\bibitem{ATLAS-CONF-2011-063}
{The ATLAS Collaboration}, {\em {Muon reconstruction efficiency in reprocessed
  2010 LHC proton-proton collision data recorded with the ATLAS detector}\/},
  {ATLAS Note} {\bf ATLAS-CONF-2011-063} (2011)  .

\bibitem{ATLAS-CONF-2011-080}
{The ATLAS Collaboration}, {\em Reconstruction and Calibration of Missing
  Transverse Energy and Performance in W and Z Events in ATLAS Proton-Proton
  Collisions at $\sqrt{s}$=7 TeV\/},  {ATLAS Note} {\bf ATLAS-CONF-2011-080}
  (2011)  .

\bibitem{Hagiwara}
K.~Hagiwara, R.~D. Peccei, and D.~Zeppenfeld, {\em Probing the weak boson
  sector in $e^+e^-\rightarrow W^+W^-$\/},  Nucl. Phys. {\bf B282} (1987)  253.

\bibitem{EllisonWudka}
J.~Ellison and J.~Wudka, {\em {Study of trilinear gauge boson couplings at the
  Tevatron collider}\/},
  \href{http://dx.doi.org/10.1146/annurev.nucl.48.1.33}{Ann. Rev. Nucl. Part.
  Sci. {\bf 48} (1998)  33},
\href{http://arxiv.org/abs/hep-ph/9804322}{{\tt arXiv:hep-ph/9804322}}.

\bibitem{mc@nlo4manual}
S.~Frixione, F.~Stoeckli, P.~Torrielli, B.~R. Webber, and C.~D. White, {\em
  {The MCaNLO 4.0 Event Generator}\/},
\href{http://arxiv.org/abs/hep-ph/1010.0819}{{\tt arXiv:hep-ph/1010.0819}}.

\bibitem{frequentist}
G.~Cowan, K.~Cranmer, E.~Gross, and O.~Vitells, {\em {Asymptotic formulae for
  likelihood-based tests of new physics}\/},
  \href{http://dx.doi.org/10.1140/epjc/s10052-011-1554-0}{Eur. Phys. J. {\bf
  C71} (2011)  1554},
\href{http://arxiv.org/abs/hep-ph/1007.1727v2}{{\tt arXiv:hep-ph/1007.1727v2}}.

\bibitem{LEPlimits}
{LEP} Collaboration, {The LEP Collaborations ALEPH, DELPHI, L3, OPAL, and the
  LEP TGC Working Group}, {\em
  {http://lepewwg.web.cern.ch/LEPEWWG/lepww/tgc/}\/},    (2005).

\bibitem{CDFlimits}
{CDF} Collaboration, T.~Aaltonen et al., {\em
  {http://www-cdf.fnal.gov/physics/ewk/2008/WZatgc/}\/},    (2008).

\bibitem{D0limits}
{D0} Collaboration, V.~M. Abazov et al., {\em {Measurement of the
  $WZ\rightarrow \ell\nu\ell\ell$ cross section and limits on anomalous triple
  gauge couplings in $p\bar{p}$ collisions at $\sqrt{s}$ = 1.96 TeV}\/},
  \href{http://dx.doi.org/10.1016/j.physletb.2010.10.047}{Phys. Lett. {\bf
  B695} (2011)  67},
\href{http://arxiv.org/abs/hep-ex/1006.0761}{{\tt arXiv:hep-ex/1006.0761}}.

\end{thebibliography}\endgroup


\end{document}